# Real-time programmable mode-locked fiber laser with a human-like algorithm

Guoqing Pu, Lilin Yi*, Li Zhang, Weisheng Hu
State Key Lab of Advanced Communication Systems and Networks, Shanghai Jiao Tong University, Shanghai, 200240, China, *lilinyi@sjtu.edu.cn

Nonlinear polarized evolution-based passively mode-locked fiber lasers with ultrafast and high peak power pulses are a powerful tool for engineering applications and research. However, their sensitivity to polarization limits their widespread application. To address this, automatic mode-locking immune to environmental disturbances is attracting more concerns. An experimental demonstration of the first real-time programmable mode-locked fiber laser is presented, which is enabled by our proposed human-like algorithm. It combines human logic with a machine's speed, computing capability, and precision. The laser automatically locks onto multiple operation regimes such as fundamental mode-locking, harmonic mode-locking, Q-switching, and even Q-switched mode-locking without physically altering its structure. The average initial mode-locking times (with randomly-assigned original polarization states and recover time from detachment under 10 successive experiments) are only 3.1 s and 58.9 ms. We believe this intelligent laser's superior performance can find practical applications in engineering and scientific research.

Ultrafast fiber lasers have extensive applications in optical frequency measurements[1-4], high-resolution atomic clocks[5,6], signal processing[7,8], ranging metrology[9,10], and astronomy[11,12]. The mode-locked fiber laser (MLFL) is the primary means of generating ultrafast pulses with extremely high peak power and an incredibly wide spectrum. Nonlinear polarization evolution (NPE) is particularly favored by researchers as the main method to realize passively MLFL. This is because of its simple structure, superior performance, and rich nonlinear dynamics[13-21]. Through flexible polarization control, NPE-based MLFL can produce harmonic mode-locking (HML)[16,17,22] with high repetition rate, Q-switching (QS)[23-25] with high pulse energy, and Q-switched mode-locking (QML) with high pulse peak power[25,26]. These operating regimes combined with the fundamental mode-locking (FML) regime, show the rich nonlinear dynamics of NPE-based MLFL. However, manual polarization control is difficult to locate in other rare regimes (except for FML regime) in passively MLFLs because the corresponding polarization solution space is much narrower. Under the disturbance of thermal instability and mechanical vibration, polarization control in NPE-based MLFL becomes more difficult[27,28]. Thus, fast and programmable polarization control based automatic mode-locking is urgently required. Automatic mode-locking through direct sweeping of polarization space has been reported with the help of electrical polarization control[18-21]. This method is straightforward but has low-efficiency because the entire polarization space is extremely large. Recently, genetic and evolutionary optimization algorithms have been used to achieve automatic mode-locking through fast polarization searching[27-31]. Automatic mode-locking applying machine learning and deep learning have also been demonstrated[32-36]. Owing to the complexity of the genetic algorithm/evolutionary algorithm/deep learning methods, the automatic mode-locking procedure is time-consuming. Off-line searching the polarization space is required in the above-mentioned methods, further limiting the automatic mode-locking time. For instance, the automatic mode-locking time is around 30 minutes[27,29] and even up to 2 hours when searching HML regimes[30]. The long mode-locking time and long recover time from disturbance limits the practical applications in engineering. It will also limit the transient pulse formation observation using the time-stretch dispersive Fourier transform (TS-DFT) for scientific research[37-39]. Therefore, there have been limited studies investigating the rich transient dynamics in NPE-based MLFL.

Genetic, evolutionary, and neural network-based deep learning algorithms are considered as the main intelligent algorithms in the artificial intelligence field. The intelligent algorithm-based MLFLs are known as autosetting laser[30], smart laser[27], and self-tuning laser[33,36] (because the lasers can automatically reach the desired operation regime). However, it cannot be regarded as an *intelligent* laser yet. In our opinion, an intelligent MLFL should have the capability of automatically locking on and switching to any possible operation regime in a fixed laser cavity such as FML, HML, QS, and QML. Additionally, the polarization searching algorithm should learn and follow human logic. None of the previous automatic MLFLs meet these criteria, where only one or two regimes have been achieved and the intelligent algorithms are completely different from human logic. Even though the intelligent algorithms can achieve the objective of mode-locking through global optimization, the searching time is too long (even longer than an experienced researcher). Compared with a machine, a human's core advantage is logic. The main advantages of machines include speed, computing capability, and precision. Hence, a genuine intelligent algorithm should combine the advantages of humans and machines.

In this study, the first real-time intelligent MLFL is experimentally demonstrated that can automatically lock on and switch to FML, HML, QS, and QML regimes using our human-like algorithm (HLA). HLA combines the advantages of humans and machines. It consists of advanced Rosenbrock searching (ARS), random collision recovery (RCR), and monitoring. The ARS algorithm guides the passively MLFL from free running to the desired operation regime. RCR is our proposed algorithm to recover the passively MLFL back to the desired operation regime from detachment induced by environmental disturbances. Both ARS and RCR learn and follow human's logic in the polarization tuning process for mode-locking. We set a series of objective functions for different operation regimes and use them as the judgment criteria during the entire process. Attributed to the HLA, the shortest mode-locking time from free running towards FML regime expends only 0.22 s. This is the fastest among all automatic mode-locking realizations to the author's knowledge. For 10 repeated experiments with randomly assigned original states of polarization (SOP), the average mode-locking time is only 3s. The pulse width varies from 115 fs to 155 fs even though the corresponding driven voltages of the electrical polarization controller (EPC) are different. This shows the repeatability of the intelligent system. The fastest recovery from vibration-induced detachment via RCR of HLA costs 14.8 ms (which is also a new record). The fast recover time from detachment can protect the subsequent devices from being damaged by the high power continuous-wave (CW) light. Note, the HLA is implemented in a field-programmable gate array (FPGA), which contributes to its simplicity. Only an analogue-to-digital converter (ADC) and digital-to-analogue converter (DAC) are required to construct the feedback-loop. Therefore, this intelligent MLFL with multi-regime operation, fast mode-locking, low cost, and high stability is suitable and ready for engineering applications. Furthermore, through recording several sets of voltages leading to different operation regimes (i.e., the experienced values) the laser can achieve microsecond-level switching among different operation regimes. Thus, we believe this intelligent MLFL will also be a powerful tool to observe and study the transient dynamics among different operating regimes assisted by the TS-DFT technique[37-39]. Additionally, this intelligent control system could be used in any feedback system with complex nonlinear dynamics but without a known nonlinear model.

**Results**

**Human-like algorithm.** Figure 1a compares recent automatic mode-locking works and our work on initial lock time, recover time, and number of regimes. The initial lock time of our laser is 400 times shorter than the next shortest initial lock time to date. Our laser's recover time is 2000 times shorter than the previously reported shortest recover time. Additionally, our laser can realize 5 different regimes in an identical setup (the most among these works). The results of comparisons illustrate the efficiencies of both the proposed HLA and real-time experimental setup.

HLA mainly consists of three portions: ARS algorithm, RCR algorithm, and discrimination of each regime. HLA (Fig. 1b) starts from ARS, which is input into the discrimination-based monitoring phase after locking onto the desired regime. HLA then focuses on monitoring (unless detachment is detected). RCR attempts to pull the laser back on track afterward. HLA goes back to the monitoring process if the recovery is successful. Otherwise, restarting ARS is necessary to anchor a new desired point. This software scheme is denoted as HLA because our proposed algorithm has many features in common with human's logic and behaviours in the manual mode-locking process. The ARS process is somewhat an analogy of human's behaviours in polarization tuning. Let us consider how a highly-experienced researcher manually tunes the polarization controller (PC) for the FML regime. Initially, three or four knobs of the PC are randomly tuned to approach the critical FML regime. Correspondingly, ARS randomly assigns the SOP of EPC at the starting point. When the researcher notices the peaks' voltages of the waveform increasing (using an oscilloscope), they continue tuning in the current direction (just like the ARS pace reward mechanism). Otherwise, the researcher tunes in the opposite direction, which corresponds to the pace punishment mechanism in ARS. Near the mode-locking regime, the tuning step switches from coarse to fine. Therefore, the ARS algorithm follows similar logic and behaviour as a human but with much faster speed and tuning precision. As most detachments are considered to be the result of environmental perturbations (with only minor SOP drift), humans usually conduct random fine tuning on the original mode-locking SOP. If the laser cannot recover from the detachment by random fine tuning, the tuning step is enlarged by the researcher using the reward and punishment mechanism. The RCR algorithm's response is similar to the human's after detachment happens but with much faster speed. Hence, we consider the HLA combines a human's logic and a machine's speed, computing capability, and precision.

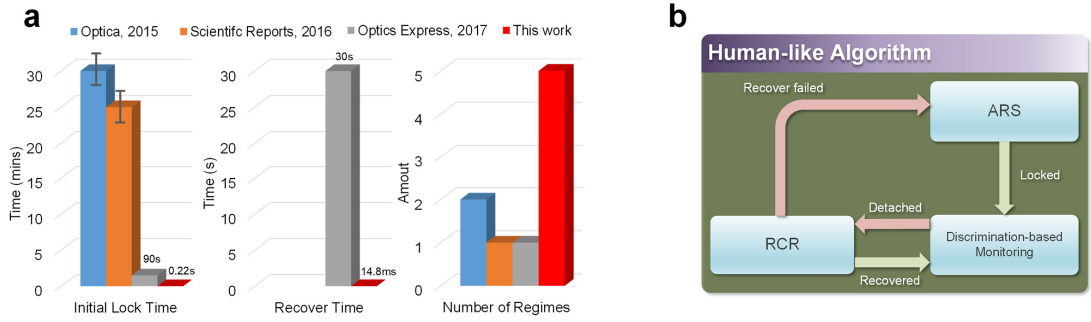

**Fig. 1** Comparisons with other recent works and HLA. **a** Comparisons between recent automatic mode-locking works and our work on initial lock time, recover time, and number of regimes. **b** Schematic of HLA. ARS keeps on running until the desired regime is achieved. A discrimination-based monitoring phase is used to detect detachment. Once detachment appears, RCR tries to return to the desired regime. HLA initializes ARS from the breakpoint when RCR fails to recover.

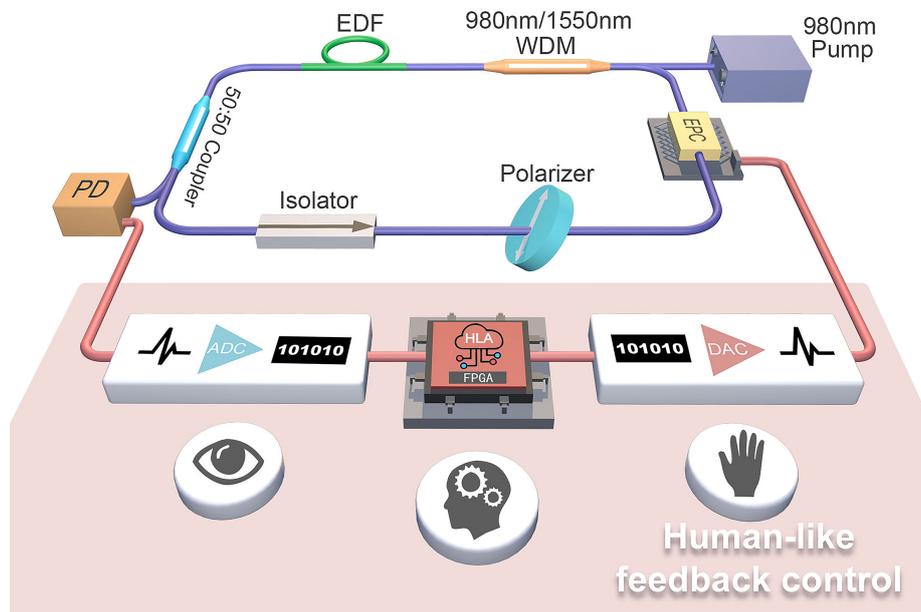

**Fig. 2** Experimental setup. A 980 nm laser pumps an Erbium-doped fiber (EDF) as the gain medium. A 50:50 optical coupler keeps half of the power inside the cavity and the other half is for detection. The isolator guarantees unidirectional optical transmission. The polarizer is the key component in NPE-based mode-locking. The EPC is utilized for polarization tuning. A photodetector (PD) is used to complete the conversion from optical to electrical. ADC, FPGA, and DAC form the human-like feedback panel.

**Experimental setup.** Figure 2 illuminates the NPE-based passive MLFL, with a cavity length of ~28.7 m (see Methods for details). The core optical component in the cavity is an EPC. Through DC voltage adjustment, arbitrary SOP can be obtained by the EPC. In the electronic part, ADC, FPGA, and DAC together constitute the human-like feedback control panel. The ADC functions like a human's eyes in monitoring the waveforms. The sampled waveform is then delivered to FPGA (the computing centre which performs a similar function to the human's brain in this instance) for the desired regime searching utilizing the ARS algorithm. During the searching process, FPGA adjusts DC voltage through a DAC. DAC translates the instructions from FPGA to DC voltage and finally acts on the EPC. This process is quite like a human's hand tuning a PC or wave plate. These steps are repeated one by one until the desired regime is achieved.

**Advanced Rosenbrock searching and random collision recovery.** The ARS algorithm is the core component of the HLA and is illustrated in Fig. 3a (see Methods for details). It is based on the traditional Rosenbrock algorithm, which is an unconstrained direct search method[40]. The distinguishing feature of ARS from the traditional Rosenbrock algorithm is a unique exit mechanism named *patience*. Patience involves a preset parameter which is the maximum of successive exploration failures that the ARS algorithm can tolerate. Once patience is exceeded, the current optimization is terminated immediately. Subsequently, a new exploration starts from a new random point. Due to this unique exiting mechanism, the ARS algorithm outperforms the traditional Rosenbrock algorithm in exploiting the potential towards various regimes of each initial point and avoids the difficulty of moderately choosing the exiting variation threshold (as required in the traditional Rosenbrock algorithm). In this experiment, DC voltages of EPC are the variables and the size of the voltage space is extremely large at $4096^4$ (as confined by the resolution of DAC). The objective of ARS is chosen to be some special feature of the time domain waveform or fast Fourier transform (FFT) result depending on which regime the laser is desired to operate on.

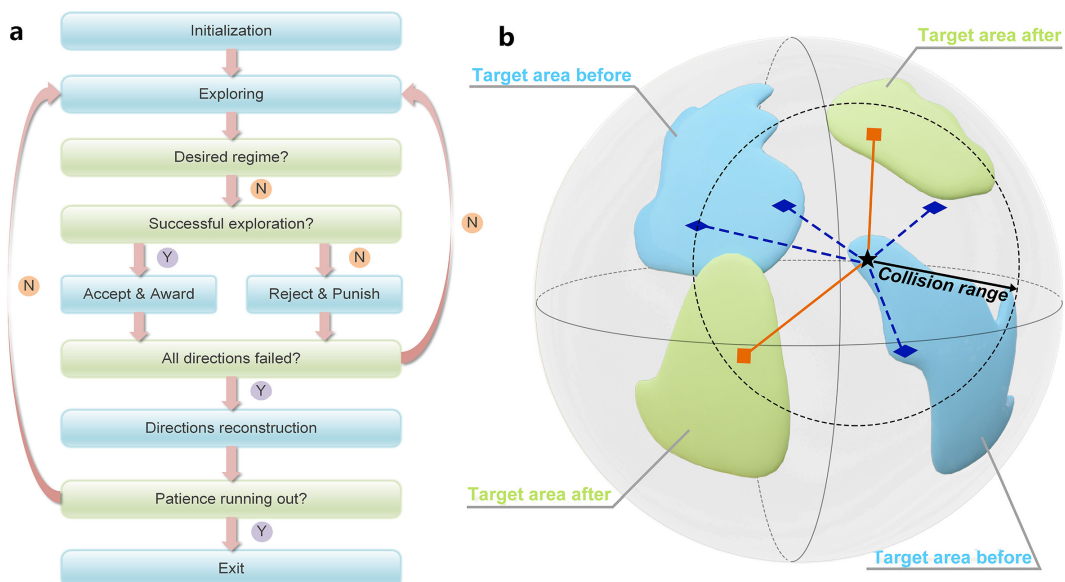

**Fig. 3** ARS and RCR. **a** The flowchart of ARS. The first step is the initialization of necessary parameters. Then exploration is executed on each direction, round by round, until the desired regime is achieved. The exploration is successful when the new objective function value outweighs the old one. Otherwise, it is considered a failed exploration. After a thoroughly failed round the exploration of each direction fails. Gram-Schmidt orthogonalization is then used to construct the new directions for next-stage exploration. ARS exits when the variable variation is too small or patience (the maximum number of successive exploration failures tolerated) is running out. **b** RCR principle. Target area, which corresponds to the desired regime, before and after the birefringence variation are illustrated on the Poincaré sphere via different colors. The black star is the original target point and the dashed-line circle expresses collision range of RCR. The blue diamonds fail to recover and the red squares successfully recover back to the desired regime through RCR.

SOP of light in fiber always varies with the birefringence variation induced by temperature, strain, mechanical vibration, and other environmental factors. SOP sensitivity to environmental disturbance has frustrated MLFL development for a long time because it induces frequent detachments, increases the instability of MLFL, and simultaneously decreases its practicality in many scenarios. Fortunately, this kind of SOP variation drifts slowly when external factors change slowly[41]. Therefore, after detachment from the desired regime, current SOP is close to the altered SOP area corresponding to the desired regime[28]. RCR principle is illuminated in Fig. 3b, target area, which corresponds to the desired regime, before and after the birefringence variation are illustrated on the Poincaré sphere via different colors. The black star is the original target point, but fails to lock after birefringence variation. Inside the dashed-line circle is the area that the RCR algorithm operates on, which is defined by collision range (a preset parameter in RCR). Note, the collision range is smaller than the tuning step in the ARS algorithm. The blue diamonds are failed attempts of RCR and the red squares successfully recover back to the desired regime via RCR. The RCR algorithm changes current SOP by adding a tiny SOP variation. Next, the RCR algorithm directly discriminates on the waveforms under the altered SOP until the desired regime is detected. Nevertheless, RCR is judged as a failure after dozens of trials have been conducted in which case the algorithm will reboot the ARS process from the breakpoint.

**Discrimination.** Pulse count proves to be an effective method in discriminating mode-locking regimes[18-20]. To eliminate higher HML regimes, the dual-region count (DRC) scheme is proposed. DRC is illustrated in Fig. 4a, blue pulses in green regions are desired while red pulses out of green regions are considered to be noise. Threshold 1 (blue dashed line in Fig. 4a) is built for the desired pulse count. Threshold 2 (red dashed line in Fig. 4a) is a limit that noise should not exceed. Ideally, the count in green regions (denoted as $C_1$) should satisfy:

$$C_1 = \frac{N_{total\_pts}}{N_{period\_pts}} \pm 1 \qquad (1)$$

whereas the count out of green regions $C_2$ should be zero.

For real-time automatic mode locking, successful and fast discrimination of different operation regimes is crucial. A set of objective functions are designed to address this problem. The waveform is evaluated as the FML regime when DRC is satisfied. Our previous criterion[18], which involve setting a threshold on the variation of pulse train, is unable to execute because of the ADC sampling rate constraint. As is well known, mode locked pulses have a considerably larger amplitude than the free running state. The objective function of the FML regime derived from this feature is:

$$O_{FML} = \frac{1}{m}\sum_{i=1}^{m} A_i \qquad (2)$$

where $A_i$ denotes the amplitude of pulse and $m$ is the pulse count from DRC. Hence, the objective function of FML is the average of counted pulses' amplitudes.

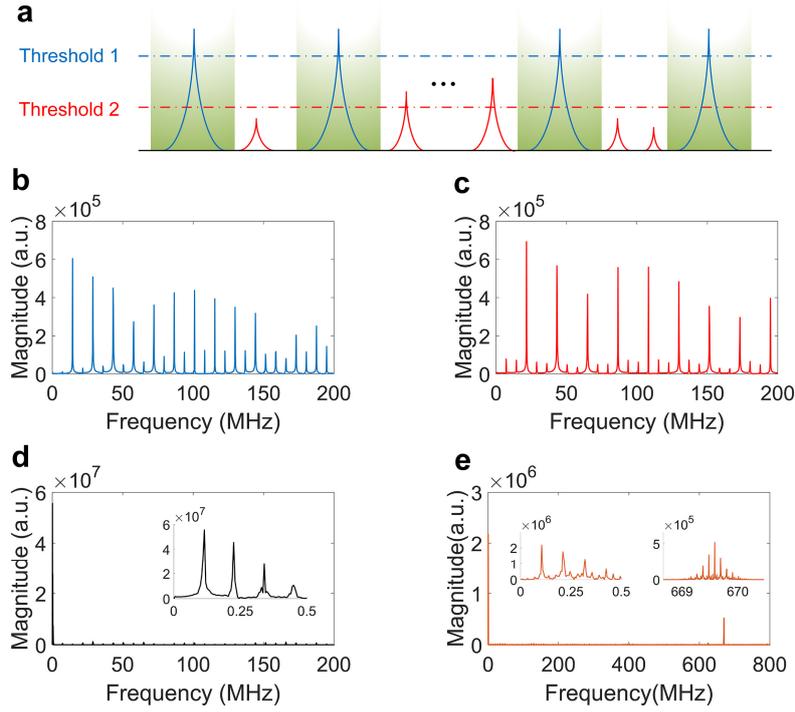

**Fig. 4** DRC scheme and FFT results. **a** DRC scheme. DRC is a crucial part in discriminating mode-locking regimes. Blue pulses in green regions are desired while red pulses are noise. Threshold 1 (blue dashed line) is set for the desired pulse count and threshold 2 (red dashed line) is a restriction for noise. **b** FFT result of the second-order HML regime. Below 200 MHz, the second spectral line is the strongest and all even spectral lines are much stronger than the odd spectral lines. **c** FFT result of the third-order HML regime. The third spectral line is the largest and all integer multiple of three spectral lines are stronger than other spectral lines (below 200 MHz). **d** FFT result of QS regime. Almost all FFT spectral components concentrate on the low-frequency domain. (Inset) The low-frequency domain of QS regime FFT result. **e** FFT result of QML regime. Though the majority of FFT spectral components concentrate on the low-frequency domain, there are still higher-frequency components. (Left inset) The low-frequency domain of QML regime FFT result. (Right inset) The higher-frequency domain of QML regime FFT result. The center peak represents the carrier and the side peaks represent envelope.

The n-th HML regime discrimination is slightly more complicated. First, DRC satisfaction is essential. Through observation of fast Fourier transfer (FFT) results of HML regimes, it is found that the amplitude of the n-th spectral component is largest among all the spectral lines for the n-th HML regime. Furthermore, the kn-th (k >= 1 and k cannot be too large) spectral line is much stronger than other spectral lines, especially between the kn-th spectral line and (k+1)n-th spectral line. For instance, Fig. 4b shows the FFT result of the second-order HML regime below 200 MHz. It is clear that the second spectral line is the strongest and all even spectral lines are stronger than the odd spectral lines. This phenomenon is more evident in the FFT result of the third HML regime, as shown in Fig. 4c. The third spectral line is the strongest and all integer multiple of three spectral lines are distinctly stronger than other spectral lines. This relation offers the other accurate discrimination criterion for the n-th HML regime, which is that the n-th spectral line of FFT result should be the strongest. The waveform is only judged as the n-th HML regime when both time domain DRC and FFT condition are satisfied simultaneously. The objective function of the n-th HML is as follows:

$$O_{HML} = \frac{2L_n + L_{2n} + L_{3n}}{\sum_{i=1}^{M} L_i} \quad (3)$$

in consideration of special FFT characteristic of HML. $L_n$, $L_{2n}$, $L_{3n}$ represent the amplitudes of the n-th, 2n-th, and 3n-th spectral lines, respectively. The denominator of Equation (3) represents the amplitude sum of all FFT spectral lines.

Analogously, QS and QML regime discrimination can be achieved in virtue of FFT results. The majority of FFT spectral components of QS and QML concentrate on low frequencies, as illustrated in Fig. 4d and Fig. 4e. This is because of the low repetition rate of QS and the envelope of QML. The difference with the QML regime is that there are mode-locked pulses within the envelope. This indicates the existence of considerable higher-frequency spectral components. As shown in the right inset of Fig. 4e, the centre peak is the carrier frequency whereas the side peaks indicate that the carrier is modulated by a low-frequency envelope. The objective functions of QS and QML for optimization are as below, where $F_{lf}$ denotes the amplitude of low-frequency FFT spectral components and the denominator of this equation represents the amplitude sum of all FFT spectral components.

$$O_{QS\&QML} = \frac{\sum F_{lf}}{\sum_{i=1}^{N} F_i} \quad (4)$$

After the objective function value surpasses the preset threshold, the FFT result is scanned to inspect whether considerable higher-frequency spectral components exist. When the FFT result only exhibits very strong low-frequency spectral components, the current waveform is judged to lie in the QS regime. Otherwise, it is judged to lie in the QML regime.

**Operation regimes.** Under the guidance of HLA, the laser can automatically achieve multiple operation regimes without any physical structure alteration to the setup. The measured oscilloscope traces, optical/frequency spectra for FML, 2nd- and 3rd-order HML, QS, and QML under 600 mW pump power are illustrated in Fig. 5. Clearly, a fundamental repetition frequency of

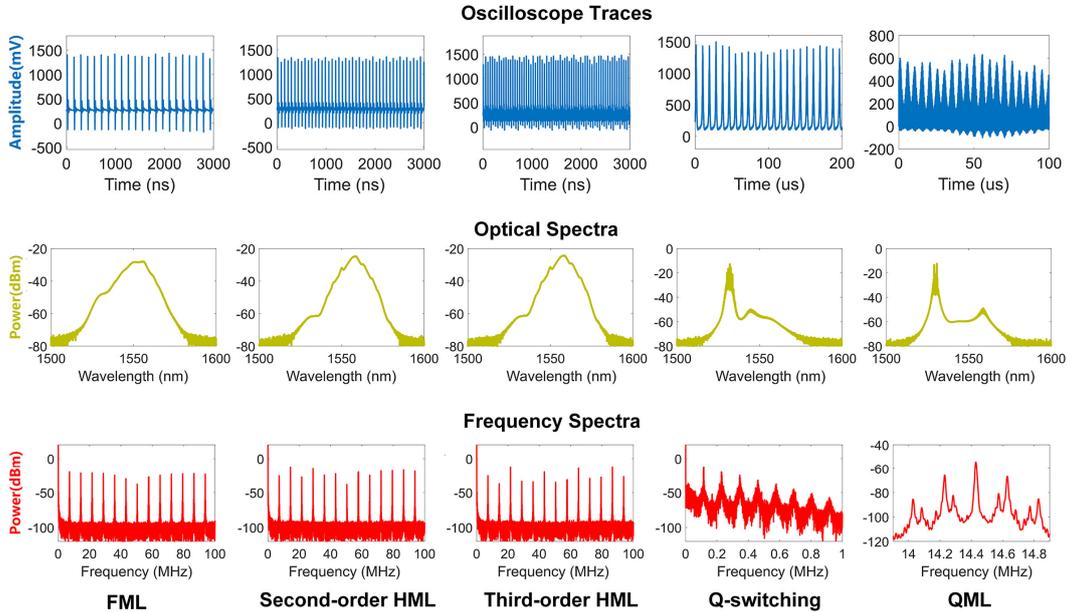

**Fig. 5** Operation regimes. From left to right: FML, second-order HML, third-order HML, QS, and QML operation regimes are shown. In each column, the top row shows oscilloscope traces, the middle row illustrates optical spectra, and the bottom plots are frequency spectra. The fundamental repetition frequency is ~7.2 MHz which is evident from the frequency spectra of the FML regime. The oscilloscope traces of two HML regimes are distinctly denser than that of FML regime and the features of their frequency spectrum are in accordance with the HML discrimination. The repetition frequency of the QS regime is 115 kHz. The envelope frequency of QML is 200 kHz and the carrier frequency is ~14.4 MHz.

~7.2 MHz is revealed by frequency spectra. From the frequency spectrum of the second-order HML regime, the second spectral line is the strongest. However, the difference between the second spectral line and the fourth line is as small as 1.7 dB. The fourth, sixth, and even eighth spectral lines are stronger than the third, fifth, and seventh spectral lines, respectively. While in the third-order HML regime, it is clear that the triple order of fundamental spectral lines is stronger than the rest of the spectral lines. Overall, the characteristics of the frequency spectra validate the effectiveness of the proposed HML objective function.

Through the frequency spectrum of the QS regime, the repetition rate of the QS pulse is determined to be approximately 115 kHz. This is far lower than the mode-locked regimes. As expected, the optical spectra width of the QS regime is much narrower compared with the mode-locked regimes because of much wider pulses. The envelope frequency of QML is 200 kHz and the basis of the carrier is the second-order of fundamental repetition frequency. Note, the laser can achieve microsecond-level switching between multiple operation regimes by recording and setting the experienced values.

## Discussion

To prove the validity of the proposed FML objective function, a scanning experiment is conducted. During scanning, the voltages of channel 1 and channel 2 of EPC are fixed. Next, the voltages of channel 3 and channel 4 of EPC are swept from 0 to 5000 mV (with a step of 24.4 mV). The scanning results of the FML objective function value and the mode-locking status are illustrated in Fig. 6a and Fig. 6b, respectively. Interestingly, the objective function value distribution presents a stair-like shape. The zone of higher objective function values is flat because the amplitude of pulse is even larger than the conversion range of ADC. The higher objective function values assemble in the zone where the voltage of channel 3 surpasses ~3400 mV. The same situation occurs with the mode-locking distribution. It is clear that almost all mode-locking points aggregate in the area where the voltage of channel 3 is over ~3400 mV. However, when the voltage of channel 3 is less than ~3400 mV, there is bare of mode-locking points. Clearly, the zone of higher objective function values and the mode-locking zone overlap. In other words, a larger average of counted pulses' amplitudes more likely correspond to the FML regime. This result is consistent with our proposed objective function. Moreover, Fig. 6c records the optimization path during FML searching. The ARS algorithm can reach the FML regime within serval effective optimization times, which reveals the high efficiency of the ARS algorithm.

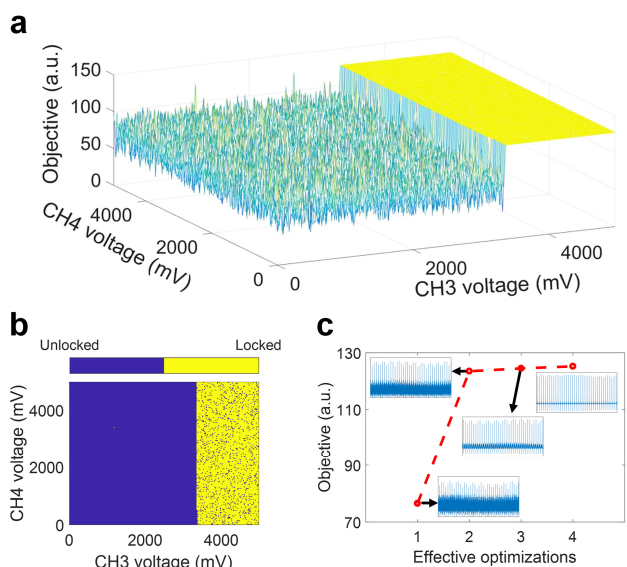

**Fig. 6** Scanning results and optimization path toward FML regime. **a** Distribution of FML objective function value. **b** Distribution of FML regime. It is clear that the zone of higher objective function values and the mode-locking zone overlap. This explains the validity of proposed FML objective function. **c** Optimization path toward FML regime. ARS finds the FML regime in serval effective optimization times. (Insets) oscilloscope traces of each effective optimization on FML objective function value.

To demonstrate the time-consumption performance of our system, 10 successive experiments of FML on initial mode-locking and recovery from detachment are performed. The results are shown in Fig. 7a. The initial DC voltages of EPC for each time ARS is completely random correspond to random original SOP in the cavity. It only takes a few seconds from free running to the desired FML regime because of the high-efficiency ARS algorithm and valid objective function. The mean of 10 initial lock times is only 3.1 s (the longest is 12.8 s). The shortest cold boost only takes 0.22 s. To the best of the author's knowledge, this is the fastest time recorded for automatic mode-locking. For recover time, 10 successive recoveries all finish within 100 ms. The detachments originate from bending or twisting the fiber manually. The longest recover time is 95.2 ms and the mean of 10 recover times is 58.9 ms. The shortest recover time is 14.8 ms. Still, it is the fastest recorded recover time (to the author's knowledge). Except for the three recoveries marked in grey, the other seven recoveries are successfully achieved through the proposed RCR algorithm (and with less time than the three relocking events via ARS). The longest time consumption of the three relocking via ARS is 95.2 ms, which is much less than the initial mode-locking time through ARS. The reason for this is that the distance between the current SOP and the desired SOP area is out of the collision range in the RCR algorithm. However, the current SOP is not far from the desired SOP area (due to the slow drift nature of the SOP under environmental disturbance) so it is easily reached by ARS with a larger tuning step than RCR. Owing to the extremely fast recovery, the detachments from mode-locking are visually undetectable and only the software can record these events. This demonstrates the speed and efficiency of the RCR algorithm. The overall time consumption of ARS and RCR consists of the time for data reading from ADC, time for controlling voltage output through DAC, and time for the algorithm execution. The time consumption of the ARS algorithm is largely determined by the distance between the initial SOP and the desired SOP in the mode-locking area. However, considerable time is consumed in data reading from ADC in the RCR process. Each trial in RCR costs ~1 ms in which the response time of EPC is 10 μs at most. However, the data reading process needs ~450 μs, which is nearly half the time of each trial. The time consumed controlling voltages output via DAC is less than 2 μs and the remainder is expended on random SOP generation and discrimination. The time consumption could be further reduced with a faster ADC and a more powerful computing centre. Note, the fast recover time from detachments is extremely important for industrial applications because the optical devices followed with the MLFL are vulnerable to sustained high CW power. Meanwhile, the fast initial mode-locking time is important for observing and investigating the pulse formation process of the NPE-based MLFL. For comparison with the genetic algorithm, an offline measurement is performed using the ARS algorithm and the genetic algorithm with identical physical setup. To achieve the FML regime, the genetic algorithm with population size of 40 requires 20 minutes while the ARS algorithm only needs 1 minute. This comparison further validates the simplicity and effectiveness of the proposed ARS algorithm.

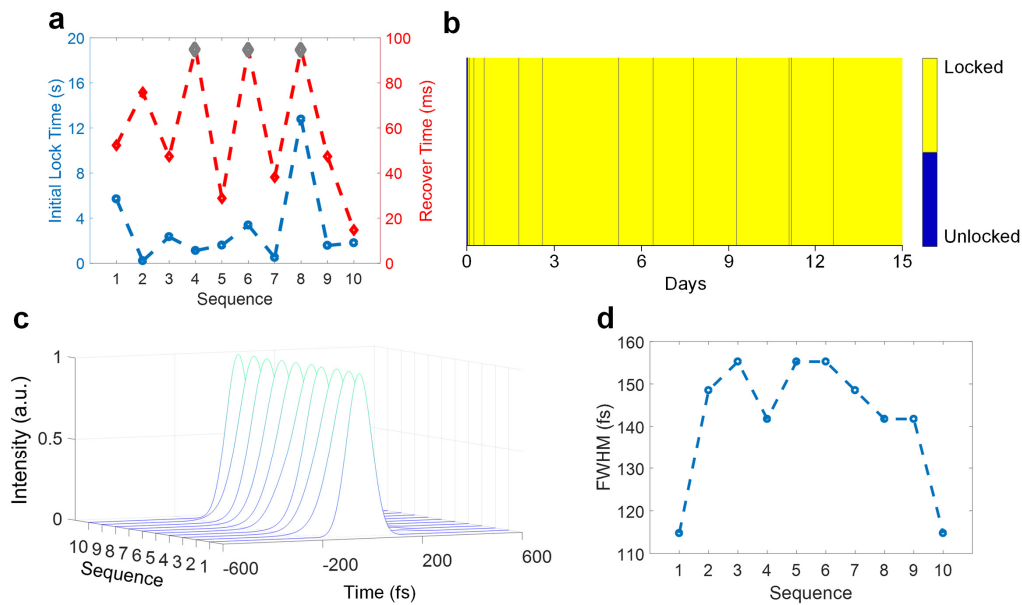

**Fig. 7** Time-consumption performance, long period running record, autocorrelation traces, and FWHM variation. **a** Time consumption of 10 successive experiments of FML on initial mode-locking and recovery. The blue line represents the initial lock time records and the shortest one is 0.22 s. The red line denotes the recovery time records and the shortest one is 14.8 ms. Except for the three gray points on red line, the remaining recoveries are all via the proposed RCR algorithm. **b** Long period running record. The initial mode-locking process costs 0.5 s and 12 detachments happen during 15 days. **c** The autocorrelation traces of 10 successive experiments. **d** FWHM variation of 10 successive experiments.

A 15-day running test was conducted under the guidance of HLA. The running record is shown in Fig. 7b. The laser reaches FML regime with the help of ARS after only ~0.5 s. There were 12 detachments in 15 days. Of those, 11 were successfully recovered using the RCR algorithm giving a mean recover time of only 31 ms. It should be noted that the long period running test was conducted in an open environment without any assistance from thermal-stability or vibration-protection devices. These results experimentally demonstrate the stability and anti-noise ability of our proposed HLA.

The autocorrelation traces and full-width-half-magnitude (FWHM) of FML pulses are measured. Figure 7c shows that the autocorrelation pulses have little difference in shape. As indicated in Fig. 7d, the maximal FWHM and the minimal FWHM in 10 autocorrelation pulses are 155 fs and 115 fs, respectively. Further, most pulse widths vary by 20 fs. This proves the stability of the output pulse shape and width of the proposed laser during repeated on-off operations.

In conclusion, the first real-time intelligent MLFL using the proposed HLA has been experimentally demonstrated. The laser is considered intelligent because it can be automatically locked on different operation regimes including FML, HML, QS, and QML without any physical structure alteration to the setup. Additionally, it can quickly recover from environmental disturbance induced detachments (not previously possible). The HLA combines human logic and machine speed, computing capability, and precision. In the author's opinion, this should be the key of an intelligent algorithm. The intelligent MLFL exhibits the fastest initial mode-locking and recover times of 0.22 s and 14.8 ms, respectively. This is attributed to the simplicity and effectiveness of the HLA. Moreover, excellent performance in a 15-day running test was observed in which the stability and anti-noise ability of the HLA was clearly demonstrated. This intelligent MLFL will surely find versatile applications in engineering and scientific research fields. The intelligent HLA is also applicable to other complex nonlinear systems without an established nonlinear model.

## Methods

**Experimental setup.** The NPE-based MLFL is indicated in Fig. 2. A 980 nm laser pumps an 8 m EDF as the gain medium. A 50:50 optical coupler keeps half of the power inside the cavity for oscillation while the other half is sent for detection. The isolator guarantees unidirectional running of the cavity and the polarizer is the key component in NPE based mode-locking. The EPC is driven by 4 channels of 0-5 V DC voltage and can respond in microseconds. A 10 GHz photodetector is used to complete the conversion from optical to electrical. Cavity length of the fiber laser is ~28.7 m. Hence, the corresponding fundamental repetition rate is ~7.2 MHz. Additionally, a 400 MSa/s ADC with 8-bit resolution, a Xilinx ZC702 FPGA, and a 100 MSa/s DAC with 12-bit resolution are used to form the human-like feedback panel.

**Advanced Rosenbrock searching.** Figure 3a shows the flow of ARS. For simplicity, variables in the optimization are denoted as $X_n = [x_1, x_2, x_3, \cdots, x_{n-1}, x_n]$ (n-dimensional). First, initialization includes setting optimization parameters and random assignment of $X_n$. After initialization, exploration is executed direction by direction. The $i$-th direction exploration vector is given by: $X_n^i = X_n + p_i d_i$, $d_i$, a $n \times 1$ where $p_i$ is the pace factor of $i$-th direction. The new objective function value is then measured under new variables $X_n^i$. When the objective function value becomes larger, the current exploration is considered successful and adopted thereafter, i.e. $X_n = X_n^i$. The pace factor is rewarded via multiplication with a number larger than 1: $p_i = \alpha p_i$. Otherwise, when the new objective function value is smaller than the old one, this exploration is considered a fail and refused. This results in $X_n$ staying the same and the pace factor being punished via multiplication with a negative number: $p_i = \beta p_i$. This means the current exploration direction is denied. Subsequently, the current exploration direction is reversed for the next round exploration. To ensure the convergence of optimization, the negative punishment factor $\beta$ should satisfy $\beta > -1$. Notably, mode-locking discrimination is inserted before the objective function value comparison to immediately discriminate the trace after exploration.

The exploration phase continues until the appearance of the desired regime or a thorough failed exploration round (a round in which exploration fails in each direction). Exploration failure means the new objective function value cannot exceed the original objective function value. Once a thoroughly failed exploration round appears, current directions are reconstructed through Gram-Schmidt orthogonalization. After reorientation, the algorithm inspects current status in case the exiting conditions are met.

The traditional Rosenbrock algorithm exits only when the variables' change is even smaller than the preset variation threshold after a round of explorations. This means the algorithm will exit after a thorough failed exploration round. Because the laser cavity is dynamic, the single calculation of objective function value cannot reflect the real situation of the cavity under a certain set of variables. However, calculating the average of several objective function values under the same set of variables is time consuming. Consequently, a unique exiting mechanism called 'patience' is introduced into the traditional Rosenbrock algorithm. Compared with the traditional Rosenbrock algorithm, the ARS algorithm with patience performs better in exploiting the potential towards various regimes of each initial point and avoids the difficulty of moderately choosing the exiting variation threshold.